\documentclass[comsoc]{IEEEtran}
\usepackage{caption}
\usepackage{color}
\usepackage[dvipsnames,table,xcdraw]{xcolor} 

\usepackage{listings}
\usepackage{multirow}
\usepackage[nomessages]{fp}

\usepackage{graphicx}
\graphicspath{ {images/}{biopics/} }  

\usepackage{comment}
\usepackage{subfig}
\usepackage{lipsum}
\usepackage[nointegrals]{wasysym}
\usepackage{textcomp}

\usepackage{mathtools}

\usepackage{paralist}
\renewenvironment{itemize}[1]{\begin{compactitem}#1}{\end{compactitem}}
\renewenvironment{enumerate}[1]{\begin{compactenum}#1}{\end{compactenum}}

\usepackage{hyperref}

\PassOptionsToPackage{hyphens}{url}\usepackage{hyperref}

\definecolor{linkcol}{rgb}{0,0,0.5}
\definecolor{citecol}{rgb}{0,0.5,0.3}
\definecolor{urlcol}{rgb}{0.3,0,0}
\newcommand{\PublicRepo}{\textcolor{blue}{\url{github.com/gsuareztangil/adrmw-measurement}}}

\newcommand{\family}[1]{\textsc{#1}}
\newcommand{\behavior}[1]{\emph{#1}}

\newcommand{\rev}[1]{\textcolor{black}{#1}}
\newcommand{\revb}[1]{\textcolor{black}{#1}}

\begin{document}

\title{Eight Years of Rider Measurement in the Android Malware Ecosystem\thanks{A shorter version of this paper appears in IEEE Transactions on Dependable and Secure Computing. This is the full version.}}

\author{Guillermo~Suarez-Tangil, 
        Gianluca~Stringhini \\[0.5ex]%
{King's College London}, {Boston University}}

\maketitle

\begin{abstract}
Despite the growing threat posed by Android malware, the research community 
is still lacking a comprehensive view of common behaviors and emerging trends in malware 
families active on the platform. Without such view, researchers incur 
the risk of developing systems that only detect outdated threats, missing the most recent ones. 
In this paper, we conduct the largest measurement of 
Android malware behavior to date, analyzing over 1.2 million malware
samples that belong to 1.28K families over a period of eight years (from 2010 to 2017).
We aim at understanding how Android malware has evolved over 
time, focusing on \emph{repackaging} malware.
In this type of threat different innocuous apps 
are piggybacked with a malicious payload (\emph{rider}), 
allowing inexpensive malware manufacturing.

One of the main challenges posed when studying 
repackaged malware is slicing the app to split 
benign components apart from the malicious ones.
To address this problem, we use differential analysis to 
isolate software components that are irrelevant to 
the campaign and study the behavior of malicious 
riders alone. 
Our analysis framework relies on collective repositories 
and recent advances on the systematization of intelligence 
extracted from multiple anti-virus vendors. 
We find that since its infancy in 2010, the Android malware ecosystem has
changed significantly, both in the type of malicious activity performed by malware 
  and in the level of obfuscation used to avoid
  detection. 
Finally, we discuss what our findings mean for Android malware detection research, highlighting areas that need further attention by the research community.
\rev{In particular, we show that riders of malware families evolve over time. 
This evidences important experimental bias in research works levering on automated systems for family identification without considering variants.}
\end{abstract}

\section{Introduction}

The Android app ecosystem has grown considerably over the recent years, 
with over 2 million Android apps currently available on 
the Google Play official market~\cite{appBrain} and with an average 
of 28 thousand uploads per day to alternative markets such as Aptoide~\cite{aptoide}. 
The number of unwanted apps has continued to increase at a similar pace. 
\rev{For instance, Google have recently removed about 790K apps that violated the market's policies, including: fake apps, spamming apps or other malicious apps~\cite{wang2018android}. 
Alarming detection rates have also been reported in other markets. }
In early 2016 Aptoide took down up to 85\% of the apps that were 
uploaded in just one month (i.e., 743K apps) after these were deemed harmful to the users. 
The poor hygiene presented by third party markets is particularly serious
because they are heavily used in countries where the Google Play Store is
censored (e.g., China and Iran).
While generally better at identifying malware, the official Google
Play store is not immune from the threat of malware either: researchers have recently discovered the largest malware 
campaign to date on Google Play with over 36 million infected devices~\cite{judyInfections}. 

The increase in the number of malicious apps has come 
hand in hand with the proliferation 
of collective repositories sharing the latest specimens 
together with intelligence about them. Virus\-Total~\cite{virusTotal} 
and Koo\-dous~\cite{koodous} are two online services available 
to the community that allow security operators to upload samples and have them
analyzed for threat assessment. While there are extensive sets of malware available, 
most past research work focused their efforts on \emph{outdated} datasets. 
One of the most popular datasets used in the literature is the Android MalGenome 
project~\cite{malgenome} and the version extended by authors in~\cite{drebin}, 
named the Drebin dataset. While very useful as a reference point, these datasets span a 
period of time between 2010 and 2012, and might therefore not be representative of 
current threats. %
More recent approaches are starting 
to incorporate ``modern malware'' to their evaluation~\cite{allix2016androzoo,lindorfer2014andrubis,mariconti2016mamadroid} 
with insufficient understanding of (i) what type of malicious activity the malware is performing or 
(ii) how representative of the whole malware ecosystem those threats are. 
Understanding these two factors plays a key role for automated 
approaches that rely on machine learning to model the notion of harm---if 
such systems are trained on datasets that are outdated or not representative, the resulting detection systems will be ineffective in protecting users. 

Despite the need for a better understanding of current Android 
malware behavior, previous work is limited. 
The first and almost only seminal work putting Android malware in perspective 
is dated back to 2012, by Zhou and Jiang~\cite{malgenome}. In their 
work, the authors dissected and manually vetted 1,200 samples categorizing 
them into 49 families. Most of the malware reported 
(about 90\%) was so-called \emph{repackaging}, which is malware that 
piggybacks various legitimate apps with the malicious payload. 
The remaining 10\% accounts for standalone pieces of malicious software. 
In the literature, the legitimate portion of code is referred to as \emph{carrier}
and the malicious payload is known as \emph{rider}~\cite{zhou2013fast}. 
In a paper published in 2017~\cite{li2017understanding}, authors 
presented a study showing how riders are inserted into carriers. 
The scope of their work spans from 2011 to 2014 and covers 950 pairs of apps.

In this work, we aim at providing an unprecedented view of the evolution of
Android malware and its current behavior. To this end, we analyze over 
\emph{1.28} million malicious samples belonging to 1.2K families 
collected from 2010 to 2017. Unlike previous studies~\cite{malgenome},
the vast number of samples scrutinized in this work makes manual 
analysis prohibitive. Therefore, we develop tools that allow us 
to automatically analyze our dataset. 

A particularly important challenge when dealing with repackaging 
is identifying the rider part of a malware sample. 
Our intuition is that miscreants aggressively 
repackage many benign apps with the same malicious payload.
Our analysis framework works in two steps. First, it leverages recent 
advances on the systematization of informative labels obtained from 
multiple Anti-Virus (AV) vendors~\cite{sebastian2016avclass,hurier2017euphony},
to infer the family of a sample. Second, it uses differential analysis 
to remove code segments that are irrelevant to the particular malware
family, allowing us to study the behavior of the riders \emph{alone}. 
Differential analysis has successfully been applied to detect 
prepackaging in the past~\cite{suarez2014dendroid,chen2015finding}, 
however it has not been used to study the behavior of the 
riders as in this study. 

We find that riders changed their behavior considerably over time.
While in 2010 it was very common to have malware monetized by sending
premium rate text messages, nowadays only a minority of 
families exhibit that behavior, and rather exfiltrate personal
information or use other monetization tricks. We also find that the use of
obfuscation dramatically increased since the early days of
Android malware, with specimens nowadays pervasively using native
code and encryption to avoid easy analysis. 
This contrasts with the amount of legitimate apps that are 
currently obfuscated---a recent investigation shows that less 
than 25\% of apps in Google Play are obfuscated~\cite{wermke2018large}, 
while we find that over 90\% of the riders active in 2017 use advanced obfuscation. 
A consequence of this is that anti-malware systems trained on both carriers and riders and/or on older datasets might not be effective in detecting recent threats, especially when they only rely on static analysis.

To the best of our knowledge, this paper presents the largest systematic study of
malicious rider behavior in the Android app ecosystem.
Our contributions are summarized as follows: 

\begin{itemize}

\item We propose a system to extract rider behaviors from repackaged 
malware. Our system uses differential analysis on top of annotated 
control flow graphs extracted from code fragments of an app. 

\item We present a systematic study of the evolution of rider behaviors in 
the malware ecosystem. Our study measures the prevalence of 
malicious functionality across time. %

\item We analyze the most important findings of our study with 
respect to the most relevant works in the area of Android malware 
detection.

\end{itemize}

\noindent
To enable replication and maintain an updated understanding of 
Android malware as time passes, we make our analysis tool publicly 
available at \PublicRepo. 
We encourage readers to visit this repository and the extended version of this paper~\cite{extended} as it provides a wider presentation of the measurements left out of this paper due to space constraints.
The dataset of samples and family labels is available at \url{http://androzoo.uni.lu}. 
\revb{
The rest of the paper is organized as follows. 
We first introduce the framework used to extract rider 
behaviors (\S\ref{sec:method}). 
Then we describe the landscape of the Android malware ecosystem (\S\ref{sec:pre}). 
We then analyze the riders alone (\S\ref{sec:rider}) 
and discuss our findings (\S\ref{sec:discussion}). 
Finally, we present the related work (\S\ref{sec:related}) and our conclusions (\S\ref{sec:conclusions}). }

\section{\rev{Methodology}}
\label{sec:method}
\subsection{Overview}
\label{sec:pre:overview}

\begin{figure}[t]
\centering
\includegraphics[width=1\columnwidth, trim=0 0 60 0, clip]{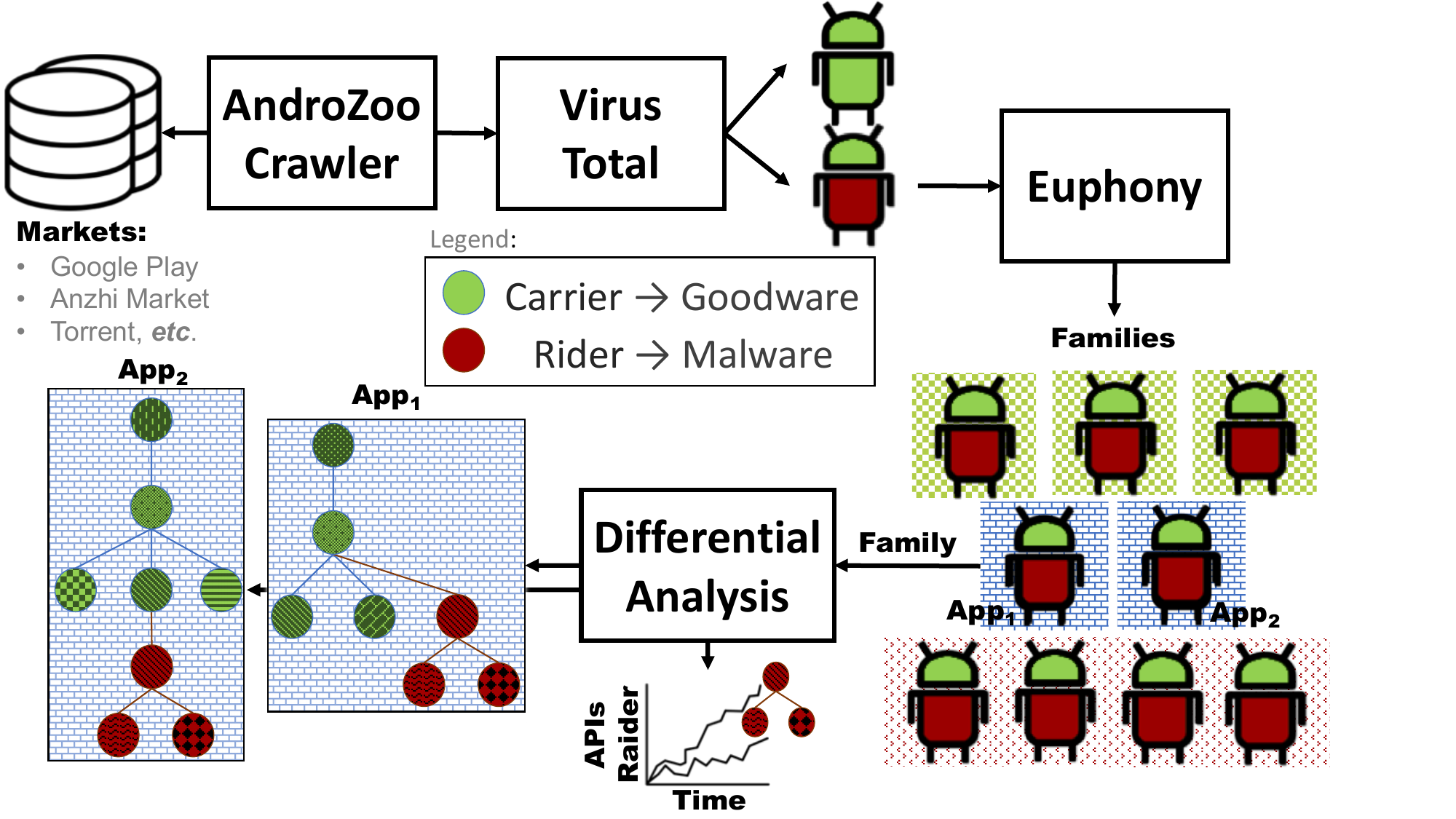}
\caption{Measurement methodology: irrelevant components are 
  removed to study the behavior of riders alone.
  }
\label{fig:pipeline}
\end{figure}

A general overview of our measurement methodology is 
depicted in Figure~\ref{fig:pipeline}.
For the sample collection we queried Andro\-Zoo in April 2017~\cite{allix2016androzoo}, 
an online repository of samples that are crawled from a variety of sources 
including Google Play, several unofficial markets, and different 
torrent sources. At the time of writing, Andro\-Zoo contains over 5.7M samples, with the 
largest source of apps being Google Play (with 73\% of the apps), 
followed by Anzhi (with 13\%). Out of all apps, %
a large portion of samples have been reported as malicious 
by different independent AV vendors (over 25\%). 
Given that Andro\-Zoo crawls apps over time and covers several markets, we
believe that this dataset is representative of the Android malware samples that
appeared in the wild. A shortcoming of this dataset is, however, that we do not
have information on how many users installed each app, and this prevents us from
estimating the population affected by such threats.
Interestingly, 
Andro\-Zoo has reported peaks of about 22\% infection rates 
in the Google Play~\cite{allix2016androzoo}, constituting the 
absolute largest source of malware. In our current snapshot 
of the Andro\-Zoo dataset, about 14\% of the apps from Google 
Play have been flagged as malware.%

The information about the AV vendors is offered by VirusTotal, 
a subsidiary of Google that runs multiple AV engines and offers 
an unbiased access to resulting reports~\cite{virusTotal}. AV 
detection engines are limited, and they certainly do not account 
for all the malware existing in the wild. This type of malware is 
known as zero-day malware and its study is out of 
the scope of this measurement. Nevertheless, both AndroZoo 
and VirusTotal keep track of the date where a sample was \textit{first 
seen} and we used this information to understand the time when 
the malware was operating. 
In addition, AV software is likely to catch up on unknown malware as time
passes, and therefore the threat of zero day malware is mitigated by the length
of our measurement.

For the label collection we relied on AV labels from 63 different vendors provided by VirusTotal. 
A common problem in malware labeling is that different AV vendors use different denominations for the same family~\cite{sebastian2016avclass}. 
To solve this problem, we unified these labels using Euphony~\cite{hurier2017euphony}, an open-source tool that uses fine-grained labeling to report family names for Android. 
\rev{Euphony clusters malware by looking at labels of AV reports obtained from VirusTotal, inferring their correct family names with high accuracy---with an F-measure performance of 95.5\%~\cite{hurier2017euphony}.
It is important to note that no a-priori knowledge on malware families is needed to do this. 
Furthermore, Euphony works at a fine-granular detection threshold. 
This means that it is able yield a label for families with samples containing only one AV report.
On average, the number of reports per sample is~8.
}

\subsection{Differential Analysis}
\label{sec:pre:dendroid}
We use differential analysis to systematically isolate software components that are 
irrelevant to our study. 
In its essence, this technique enables us to discard 
sets of observations that do not consistently appear 
in a population. In our domain, a population is a 
family for which we extract the set of methods that appear in each sample. 
Methods that are common to the different members of 
the same family are assumed to belong to the rider 
and stored for further analysis. 
Our underlying assumption is that samples from the 
same family have the same purpose and are written 
by the same authors. As part of the repackaging 
process, riders are inserted into different apps. 
Thus, it is expected to find common code structures 
within all the malware samples in the family. 

In a nutshell, our approach follows two steps: first, we extract the Control Flow
Graph (CFG) of an app and annotate each node. Second, we identify nodes that are
common throughout the malware family and extract the rider component from it.

\begin{figure}
\centering
\includegraphics[width=.7\columnwidth, trim=185 180 110 90, clip]{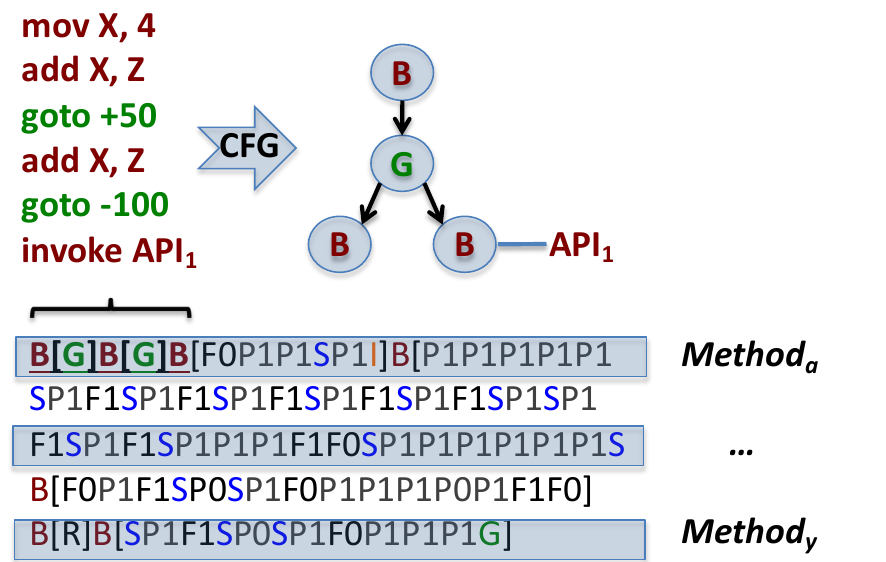}
\caption{\rev{Flattened representation of code structures in an APP.}\label{fig:acodestructure}}
\end{figure}

\noindent\textbf{CFG and graph annotation.} 
Miscreants can deliberately modify riders across 
infections to evade pattern-based recognition 
systems. To be resilient to these evasion attempts, 
we aim at obtaining an abstract representation of the code. 
The most common forms of obfuscation in Android malware 
are class and method renaming, variable encryption, 
dynamic code loading, and code hiding~\cite{suarez2017droidsieve}. 
In our work, we compute the CFG of each code fragment extracted. 
\rev{
For this, we transform the sequence of instructions seen in the binary into a list of statements defining its control flow such as blocks of consecutive instructions (namely ``basic blocks'') and bifurcations determined by ``if'' statements and jumps.
Figure~\ref{fig:acodestructure} shows an example of the code structures found in a particular app. 
Each node in the graph represents a piece of code that will be executed sequentially without any jumps. 
The CFG is then flattened based on the grammar proposed by Cesare and Xiang~\cite{cesare2010classification}.
}
We then obtain a hash (fingerprint) of each 
code structure and it is used 
to compare the set of common fingerprints for 
each family. Comparing fingerprints of smaller 
units of code to measure the similarity between 
two apps is known as fuzzy hashing. 
Fuzzy hashing has been shown to be an effective way 
of modeling repackaged apps~\cite{zhou2012detecting,
suarez2014dendroid,chen2015finding}. 
One major advantage of leveraging on a fuzzy representation 
of the CFG is an improved resistance against class and method 
renaming, as well as variable encryption. 

\rev{We then annotate each node in the CFG with the set of APIs 
(Application Program Interfaces) called in that method, to capture the semantics of each of the basic blocks in the graph. 
Annotations are simply done by adding a node to the building block where the API has been seen.
This semantics is extracted from the parameters of all Dalvik instructions 
related to \textit{invoke-*} such as \texttt{invoke-virtual}. 
These parameters typically refer to the invocation of libraries 
(including those from the Android framework related to 
reflection 
as we detail in~\S\ref{sec:rider:behavior}). 
We then parse those parameters to extract the API calls.\footnote{We also tag each API call based on the category of the library it invokes (package name) from the Android Framework as explained in~\S\ref{sec:rider:behavior}.}
This enables us to understand when certain code 
structures are using dynamic code loading and 
what is the prevalence of this behavior across 
riders. Also, the annotation of the CFG allows us 
to combine {\em fuzzy} 
hashing with a technique known as {\em feature} 
hashing~\cite{hanna2012juxtapp}. 
Feature hashing reduces the dimensionality of the data 
analyzed and, therefore, the complexity of computing 
similarities among their feature sets.}

We recursively extract fragments from all available 
resources within the app of type DEX or APK. 
The reason for this is that malware often 
hides its malicious payload in DEX or in APK files hosted as a 
resource of the main app. When the app is executed, 
the malware then dynamically loads the hidden component. 
This is referred in the literature as \emph{incognito 
apps}~\cite{suarez2017droidsieve}.

\noindent\textbf{Extraction of common methods.}
Once we have a representation of the code structures, 
we can analyze the frequency (number of apps) in which 
these methods appear across a family. In the simplest 
case, common structures will be present in all 
samples in the family. This represents the core 
functionality of the riders in this family.  
In other cases, for example when a malware family is 
composed of several subfamilies, the common code 
structures will be manifested in a subset of the 
samples. These structures are still relevant to 
understand how the family has evolved over time. 
Those methods that are not common to members of 
the same family are deemed irrelevant to 
characterizing the behavior of the app and discarded. 
Varying the percentage of common methods retained 
gives us different perspectives 
on the characterizing structures of a family. 
We next show how this is reflected in our dataset. 

\section{\revb{The Android Malware Ecosystem}}
\label{sec:pre}

\begin{figure}[t]
\centering
\includegraphics[width=1\columnwidth]{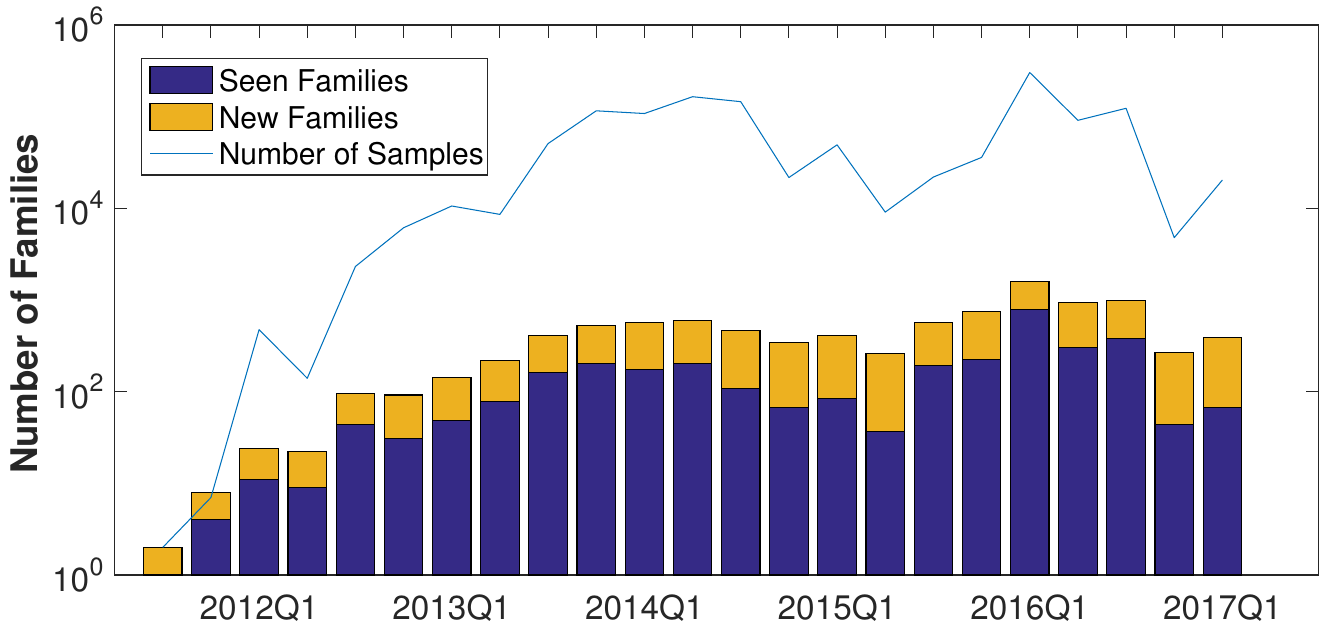}
\caption{\rev{Num. of samples and families seen per quarter.}}
\label{fig:statsNumberFamilies}
\end{figure}

\rev{Figure~\ref{fig:statsNumberFamilies} shows the number of families observed across time. 
The stacked plot distinguishes between newly observed and previously seen families at every quarter of a year. 
Seen families refer to the set of families where one specimen 
(of that family) was seen in VirusTotal prior to the referred date.}
After unifying labels and processing all samples as described 
in~\S\ref{sec:pre:dendroid}, we account for over 1.2 million apps and 1.2K families. 
The graph depicts the overall number of samples 
per quarter used in this measurement.

{\color{black} %
\subsection{Malware Family Landscape}
\label{sec:pre:top}

Our first high level analysis of our dataset aims at understanding how families evolve over time 
from a structural viewpoint. To this end, we identify the \emph{top} families 
according to the following four definitions:

\begin{itemize}
\item \textbf{Largest Families}: $top(|F_i|)$. We take a look at the 
top families ordered by the number of samples in each family ($|F_i|$), 
where $|F_i| < |F_{i-1}| \forall i=\{1, \dots, n\}$.
Note that this metric only takes into account the number of samples observed in a
family, and not the total number of installations.

\item \textbf{Prevalent Families}: $top(|Q_i^j|)$. Top prevalent families 
are ordered by the number of quarters (of a year) where a sample of a family 
was observed;  where $Q_i^j$ denotes quarter $j$ in which a sample of a 
family $i$ was seen. This metric aims to identify the longest lasting malware
families.

\item \textbf{Viral Families}: $top(|F_i|/|Q_i^j|)$, where we look at the 
ratio between how large a family is and the number of quarters in which 
the family was present. This metric aims at identifying families that
are both large and also last for a long period of time.

\item \textbf{Stealthy Families}: $top(D_i)$, where $D_i$ denotes the 
average time delta $T_{vt} - T_{dex}$ between the moment when the sample of 
a family $i$ was compiled ($T_{dex}$, as observed in the DEX file) and 
the first time the sample was seen by VirusTotal ($T_{vt}$). 
This metric looks at how difficult it is for malware detectors to 
identify the samples in a family as malicious.
\end{itemize}

Figure~\ref{fig:familiesTop} shows the distribution of apps in top 
families for each of the categories described above. The graph 
depicts the probability density of the data at different values together 
with the standard elements of a boxplot (whiskers representing the maximum 
and minimum values, and the segments inside the boxes the average and the
median). 
To cover a wider range of cases, only unique families are shown across 
all four plots. It is 
worth noting that \family{airpush} appears within the top 10 in all four categories, 
\family{leadbolt} appears in all categories except for the 
\emph{viral} one, and other families such as 
\family{jiagu}, 
\family{revmob}, 
\family{youmi}, and
\family{kuguo}  
appear in both \emph{largest} and \emph{viral} categories.

As observed in the timeline given in Figure~\ref{fig:familiesTop}, 
some families show multiple distributions indicating that there are 
outbreaks at different time periods. This is presumably when malware 
authors created a new variant of a family. %
The similar alignment for the second outbreak in some of the families 
might be explained by the latency with which AV vendors submit samples 
to VirusTotal. Also, it has been reported~\cite{graziano2015needles,wired} 
that at times miscreants use VirusTotal before distributing samples to test 
whether their specimens are detected by AVs or not. In either case, this 
is still a good indicator of how malicious behaviors span over time 
and one can observe that 2014 and 2016 reported the largest activity. 

Special emphasis should be given to \family{safekidzone} and 
\family{pirates}. The former appears as a top \emph{prevalent} 
family and the distribution of samples across time is remarkably 
uniform. This means that the miscreant has been persistently 
manufacturing new specimens across 4 years almost as if the process 
was automated. The latter starts the outbreak aggressively in mid 
2013---unlike most of the other families where infections start 
progressively.

\begin{figure*}[t] 
\centering
\subfloat[Largest Families.]{\includegraphics[width=.35\textwidth, trim = 0 5 0 0, clip]{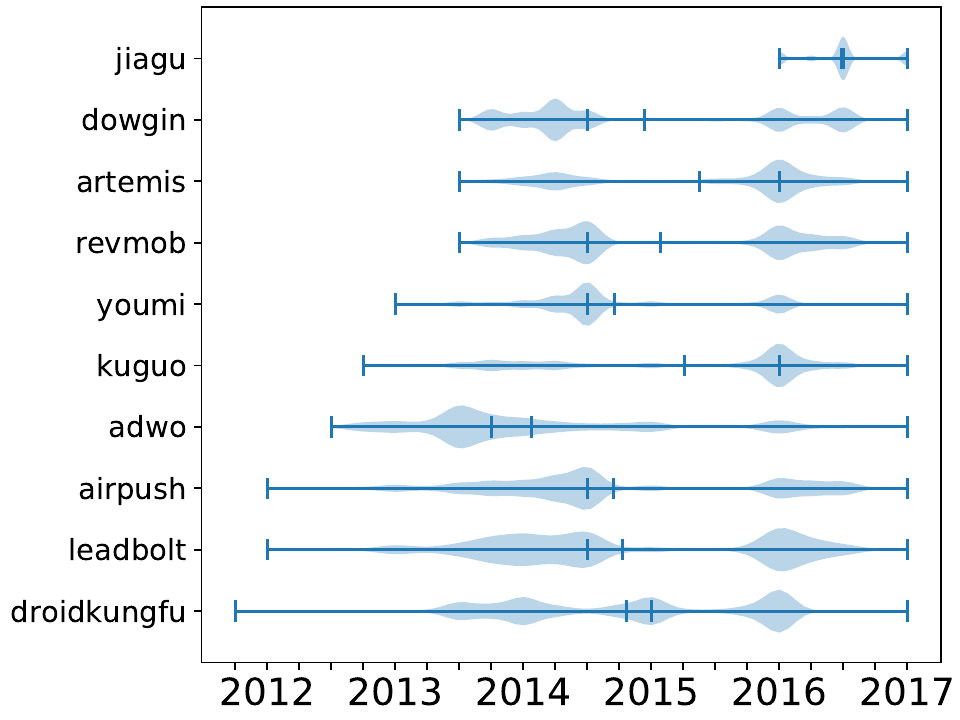}\label{fig:familiesTop:size}}
\subfloat[Prevalent Families.]{\includegraphics[width=.35\textwidth, trim = 0 5 0 0, clip]{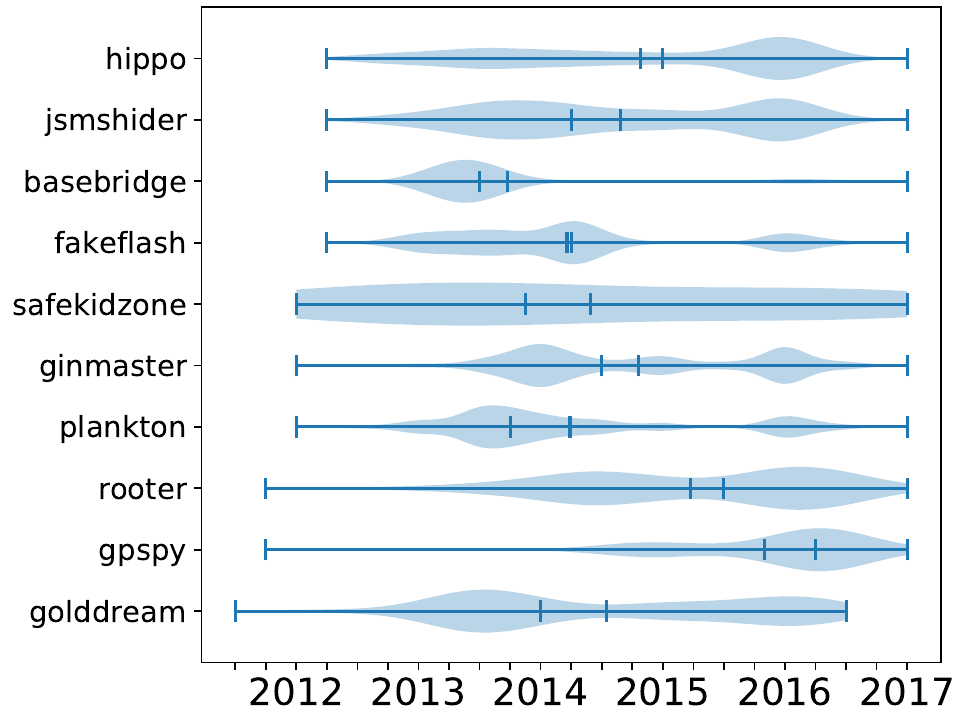}\label{fig:familiesTop:time}}
\hfill
\\
\subfloat[Viral Families.]{\includegraphics[width=.34\textwidth, trim = 0 5 0 0, clip]{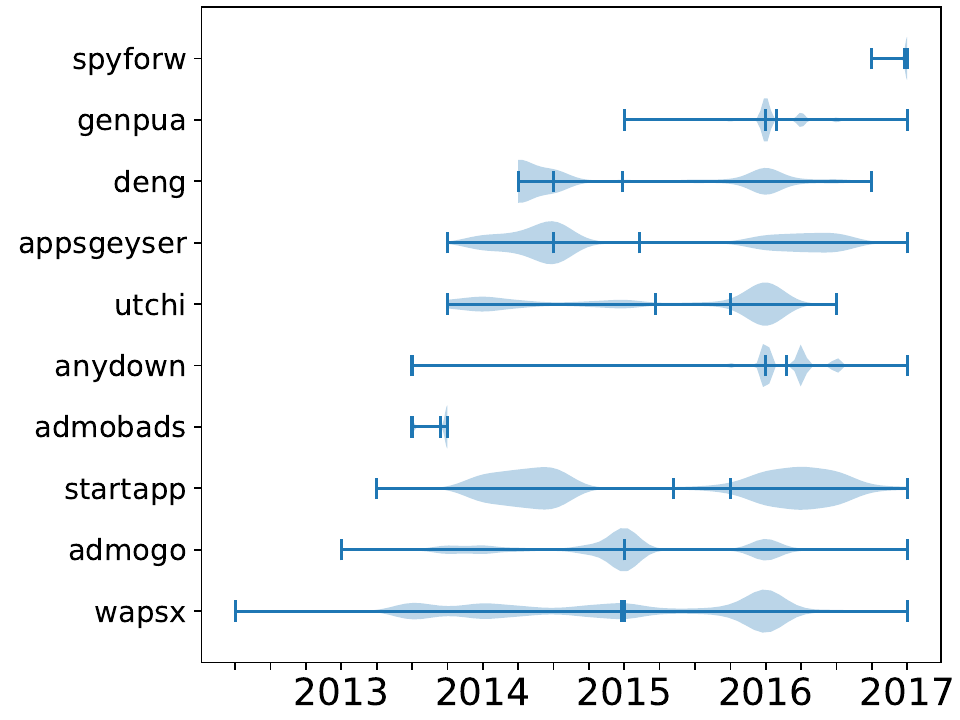}\label{fig:familiesTop:virality}}
\subfloat[Stealthy Families.]{\includegraphics[width=.35\textwidth, trim = 0 5 0 0, clip]{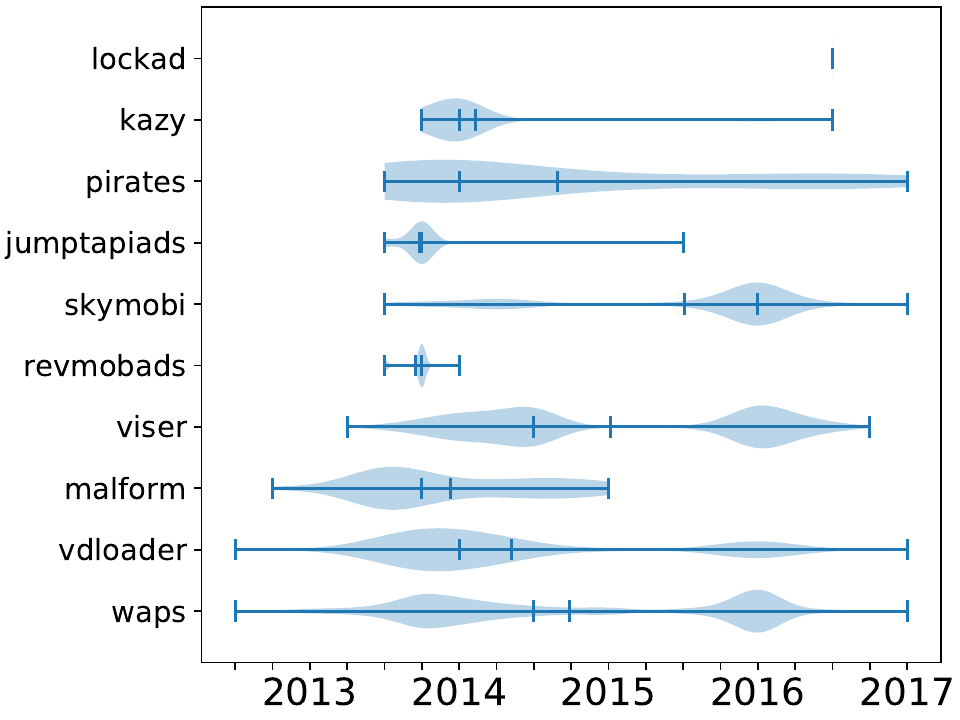}\label{fig:familiesTop:steath}}
\hfill
\caption{\rev{Distribution of samples over time for the top families in each
  category. The overall number of samples per family ranges from 29K to 262K for the largest families (a), from 143 to 11K for the prevalent ones (b), from 2K to 23K for the virals (c), and from 174 to 18K for stealthy ones (d).}}\label{fig:familiesTop}
\end{figure*}

} %

\subsection{Common Methods in our Dataset}
\label{sec:pre:common}

The total number of samples in our dataset after unifying AV labels 
accounts for almost 1.3 million apps and 3K families. Prior to 
running the differential analysis, we process the dataset to extract 
all classes and build the CFG of their methods. When processing these 
samples we found that approximately 1\% of the apps were malformed or 
could not be unpacked, leaving us 1,286,145 labeled samples. 

\begin{figure}[t]
\centering
\includegraphics[width=\columnwidth, trim=40 10 0 25, clip]{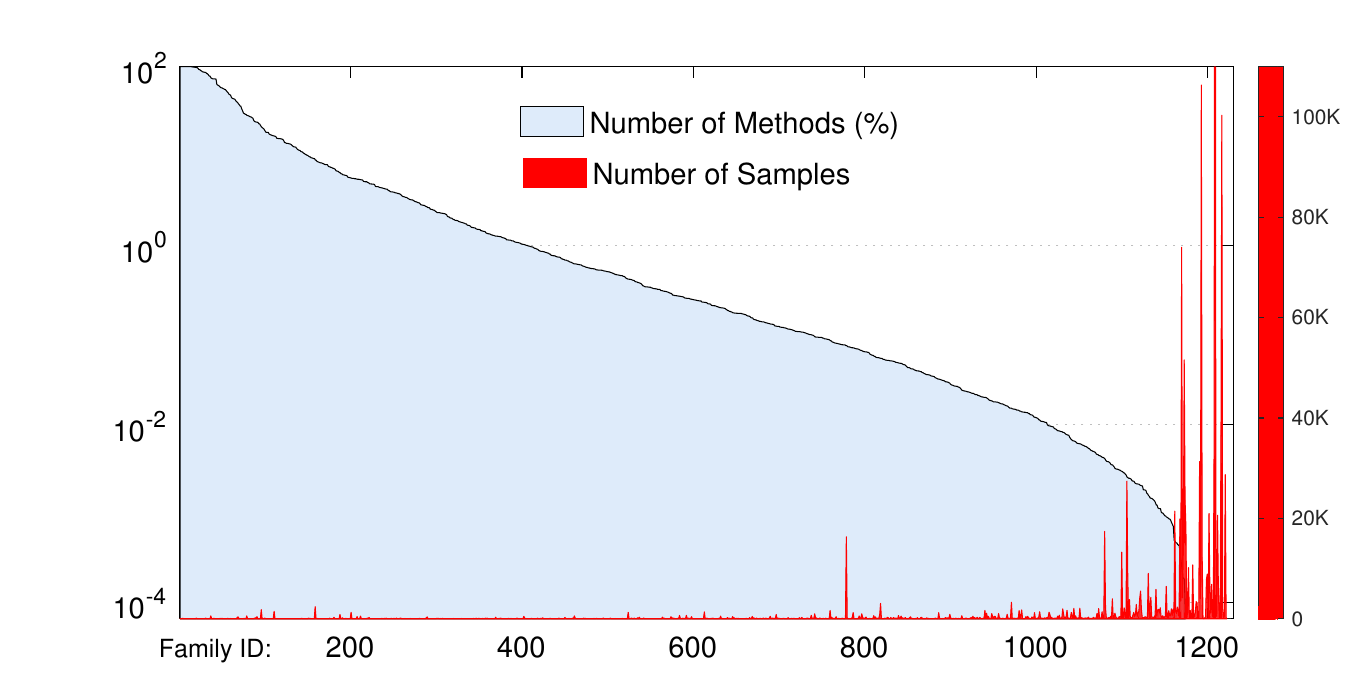}
\caption{Percentage of methods common to all samples in a family (blue area) together with the number of samples per family (red area).} 
\label{fig:commonMethodsSamples}
\end{figure}

Figure~\ref{fig:commonMethodsSamples} shows the number of methods common 
to all samples of the 1,226 families. It also displays the number of 
samples per family, which are conspicuously unbalanced. While 
most of the families have between 7 to 40 samples, there is 
one family with about 260K samples (\family{dowgin}) and there are 
two families with about 100K samples (\family{kuguo} and \family{airpush}). 
For the sake of readability the graph only displays sizes up to 100K, 
with the largest family ending off the chart. 
In general, we can observe that there are few families (with sizes 
ranging from 7 to 567 samples) where most methods are common to all apps 
in the family. In particular, there are 12 families (282 samples) 
where all of their methods appear in 100\% of the samples in the 
family. %

When all the methods seen in a family appear in all of the samples it means 
that either the family is standalone malware (without a carrier) or 
that all members in the family are repackaging the same goodware. 
We refer to the latter phenomenon as \emph{early-stage repackaging}. %
While standalone families are relevant to our analysis, there is no 
a priori way to know the type when only looking at the number of common methods. 
For this reason, we avoid running differential analysis on families where at 
least 90\% of their total methods are common to all samples of the family.
This accounts for 25 families (542 samples), which is 0.04\% of the dataset. 
As for the remaining %
families we observe that the proportion 
of methods in common varies across families regardless of their size. 
An exception to this are very large families, where the number of 
riders %
is lower than the average. 

\begin{figure*}[bt]
\centering
\includegraphics[width=.9\textwidth, trim=5 5 0 0, clip]{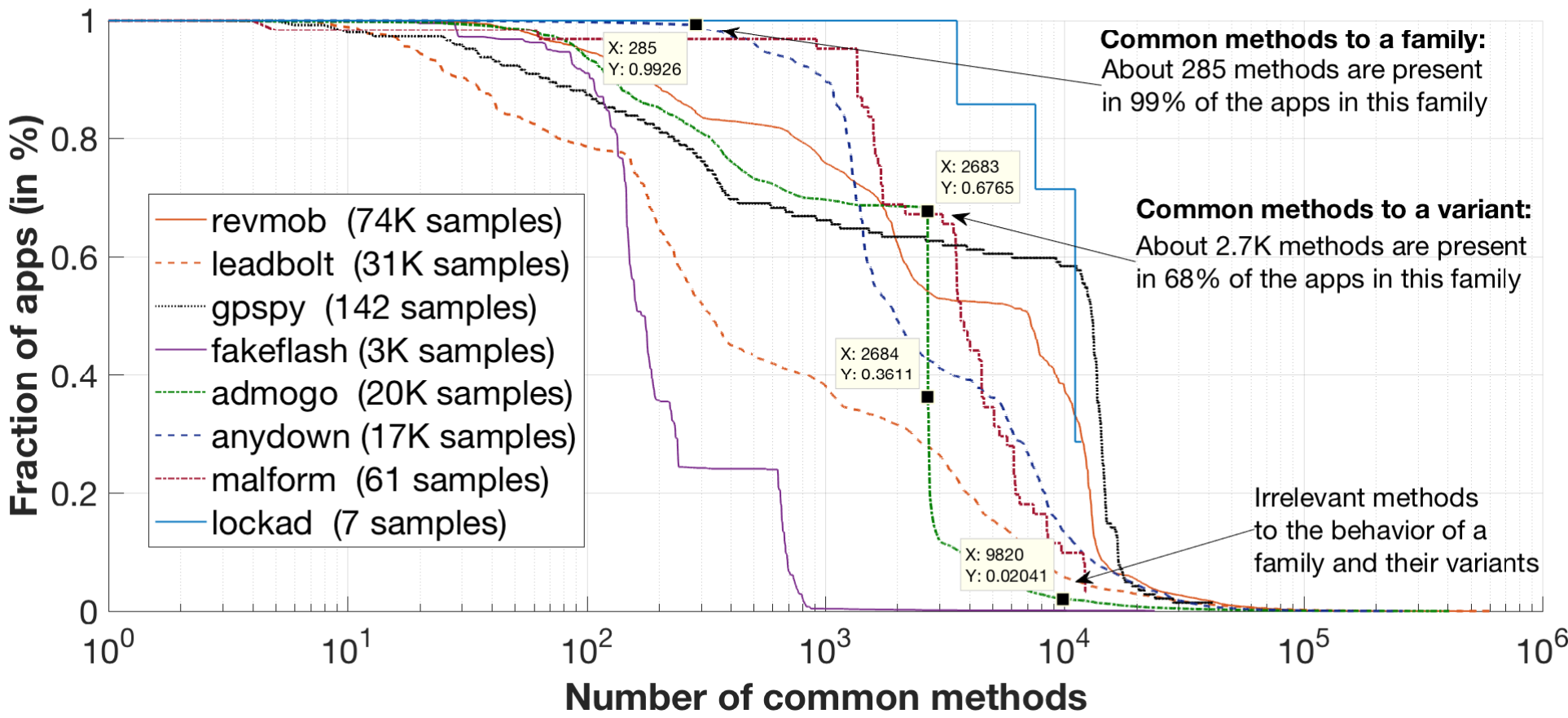}
\caption{Prevalence of methods across apps for top most popular families per category (\emph{revmob} \& \emph{leadbot} for large families, \emph{gpspy} \& \emph{fakeflash} for prevalent families, \emph{admago} \& \emph{anydown} for viral families, and \emph{malformed} \& \emph{lockad} for stealthy families).}\label{fig:statsTop10}
\end{figure*}

Malware development is a continuous process, and criminals often improve their
code producing variants of the same malware family. Our framework has the
potential to trace the appearance of such variants. 
As an illustrative example, 
we study the prevalence of methods across some of the most prevalent families. 
Figure~\ref{fig:statsTop10} shows the example of five malware families in our
dataset.
When looking at \family{anydown}, we observe that there are 285 methods common to 
99\% of the 17,000 samples in the family. The functionality embedded into 
these methods constitute the essence of the {\bf family}. 
Even when the number of methods in common to all samples of the family is small, 
there are still a number of methods common to subsets of samples from the family. 
For instance, there are only 10 methods shared by 99\% of the samples in 
\family{leadbolt}, but over 150 methods are shared by 75\% the apps (23,000).
This can be explained by the morphing nature of malware. It is commonplace to see 
malware families evolving as markets block the first set of apps in the 
campaign~\cite{googleblogVariants}. 
This ultimately translates into different {\bf variants} that are very similar. 
Interestingly, we can observe that the boundaries defining variants of a family 
are sometimes well established. This is the case of \family{admogo}, a family that 
altogether has about 20,000 samples. We can see a variant with 2,683 methods 
common to 67.65\% of the samples, and we can see another variant with one 
additional method in common (i.e., 2,684) shared by only 36.11\% of those samples (c.f. Fig.~\ref{fig:statsTop10}). 

\subsection{Choice of a cutoff}
\label{sec:pre:cutoffs}
To be able to operate, our approach needs a cutoff.
This cutoff determines the fraction of apps within a family that need to share a
method before our system considers it as being representative of that malware
family.
Ideally, to capture the behavior of a family we would look 
at common methods in all apps (100\% threshold). However, in 
practice this is not the best choice because AV vendors can 
accidentally assign wrong labels to a 
sample~\cite{deo2016prescience,miller2016reviewer,jordaney2017transcend}. 
In our experiments we set this threshold to 90\% 
based on the F-measure performance reported by Euphony 
(92.7\%\textasciitilde95.5\%~\cite{hurier2017euphony}). 
We consider this threshold to be a good value to capture 
the behavior of families, while allowing some margin for 
mislabeled samples. Note that a threshold of 90\% means 
that we look at methods that are in the interval 
$[100\%, 90\%)$. In the best case scenario, where 
there is no misclassification, we will be observing 
methods in 100\% of the apps. In the worst case, we 
will be including methods from the largest variant.

Intuitively, different cutoffs could be set to identify methods that are not
common to entire families, but are indicative of specific variants
(see~\S\ref{sec:pre:common}). Due to space constraints, we do not explore this
possibility in this paper, but in~\S\ref{sec:discussion} we discuss how this
direction could be explored in future work.
Due to the nature of the differential analysis, we can 
only study families with a given minimum number of 
samples. This number depends on the cutoff introduced. 
In particular, having a cutoff of 90\% means that 
those methods that appear in more than $\lfloor 90\% \times n \rfloor$ 
apps, where $n$ is the number of samples in a family, 
will be considered representative methods. For instance, 
in a family with $n = 3$, a representative will appear 
in at least 2 (out of 3) samples. Here, in the worst case scenario, 
the cutoff used in practice will be forced down to 66\%  
(note that $\lfloor 90\% \times 3 \rfloor = 2$ and 
$2/3 \simeq 66\%$) instead of 90\%. To avoid this, we set 
$n = 7$ (i.e.: $\lfloor 90\% \times 7 \rfloor = 6$ and 
$6/7 \simeq 85\%$). The reason we choose $n = 7$ is 
because it is close enough to the cutoff and also most 
of the families in our dataset have 7 samples or more 
(see~\S\ref{sec:pre:common}). 
Overall, we discard 0.3\% of the samples, leaving a total 
of 1,282,022 malicious apps and %
1,201 families. %

\section{Analysis of Android Riders} 
\label{sec:rider}

We use the techniques described in the previous section to study and characterize rider behaviors from 2010 to 2017. 
We first introduce the set of behaviors that we explore, 
and then give an overview of the general state.  
Finally, we study the evolution of such behaviors over time.

\subsection{Rider Behaviors}
\label{sec:rider:behavior}

To understand how malware behaves, we analyze rider 
methods from all observed families. We are primarily interested in 
learning whether malware exhibits actions related to certain 
attack goals as characterized in~\cite{suarez2014evolution}. 
In particular, we look at actions related to: 

\begin{itemize}
\item \textbf{Privacy Violations}. These actions typically 
involve queries to the Android \behavior{Content Resolver} 
framework, the use of \behavior{File Access} system, or the 
access to information such as the \behavior{Location} of the 
user, etc.%

\item \textbf{Exfiltration}. The usage of the \behavior{network} 
combined with all those actions related to privacy 
violations can indicate the leakage of personal information. 

\item \textbf{Fraud}. These actions aim at getting %
profit from the users or the services they use. 
For instance, malware can send premium rate messages 
via the \behavior{SMS Manager} or it might abuse advertisement 
networks by changing the affiliate ID to redirect revenues.

\item \textbf{Evasion}. \behavior{Hardware} serial numbers, 
versions of \behavior{firmware} and other OS configurations 
are often used to fingerprint sandboxes to evade dynamic 
analysis. 

\item \textbf{Obfuscation}. The use of obfuscation and other 
hiding techniques is a sought after technique to evade 
static analysis. Android offers options to dynamically 
load code at runtime (e.g., with \behavior{reflection}).

\item \textbf{Exploitation}. Certain apps implement technical 
exploits and attempt to gain root access after being installed. 
Most of these exploits are implemented in native code and triggered 
using bash scripts that are packed together with the app 
as a resource.%
\end{itemize}

To measure these behaviors we look at the invocation of the 
APIs used to access 
key features of the OS or data within the device. 
APIs are especially relevant in current smartphones as 
they incorporate a number of mechanisms to confine and 
limit malware activity. These mechanisms make apps 
dependent on the Android framework and all permission-protected calls 
are delivered through a well-established program interface. 
Furthermore, API calls are useful for explaining the behavior 
of an app and reporting its capabilities. 

\rev{Android APIs are organized as a collection of packages 
and sub-packages grouping related libraries together. 
On the top of the package structure we can find, for instance, 
libraries from the \texttt{an\-droid.*}, \texttt{dal\-vik.*}, and
\texttt{java.*} packages. On the next level, we can find 
sub-packages such as \texttt{android.os.*},  
\texttt{dal\-vik.\-sys\-tem.*}, or \texttt{java.\-lang.\-re\-flect.*}, among others. 
As most of the sub-packages belong to the \texttt{android.*} 
package, for the sake of simplicity, in this paper we refer to them starting 
from the second level. For instance, \texttt{android.provider.*}, 
which is a standard interface to data in the device,
is referred as \behavior{PROVIDER}. For other 
packages (e.g.: \texttt{dalvik.system.*}), we use the full name 
with an underscore (i.e., \behavior{DALVIK\_SYSTEM}).}

While program analysis can tell what are the set of API 
calls that appear in an executable, it is hard to understand 
what these calls are used for. However, there are some APIs 
that are typically used by riders for certain purposes. This is the 
case of APIs that load dynamic code, use reflection, or use 
cryptography. These are specially relevant to malware 
detection as they enable the execution of dynamic 
code~\cite{poeplau2014execute} and allow the 
deobfuscation of encrypted code~\cite{DroidNative2017}.
We summarize these functionalities as follows:

\begin{enumerate}

\item[i)]  \behavior{JAVA\_NATIVE}: This API category captures 
libraries that are used to bridge the Java runtime environment 
with the Android \emph{native} environment. The most relevant API in 
this category is \texttt{java.\-lang.\-Sys\-tem.\-load\-Li\-bra\-ry()}, which 
can load ELF executables prior to their interaction through 
the Java Native Interface (JNI). 

\item[ii)]  \behavior{DALVIK\_SYSTEM}: This category allows 
the execution of code that is not installed as part of an 
app. The following API call is key for the execution 
of \emph{incognito} Dalvik executables: 
\texttt{dal\-vik.\-sys\-tem.\-Class\-Loader.\-Dex\-Class\-Loader()}. 

\item[iii)]   \behavior{JAVA\_EXEC}: This API category allow 
apps to interface with the environment in which they are 
running. The most relevant API in this category is 
\texttt{java.lang.Runtime.exec()}, which executes 
the command specified as a parameter in a separate process. 
This can be used to run \emph{text} executables. 

\item[iv)]  \behavior{JAVA\_REFLECTION}: This category 
contains a number of APIs that make possible the 
inspection of classes and methods at runtime without 
knowing them at compilation time. This can 
be very effective to hider static analysis (e.g., 
by hiding APIs).  

\item[v)] \behavior{JAVAX\_CRYPTO}: These APIs provide 
a number of cryptographic operations that can be used 
to obfuscate and de-obfuscate payloads.

\end{enumerate}

It is important to highlight that the categories described above 
are not comprehensive and the same set of APIs can be used for 
different purposes. For instance, accessing the contacts 
(via the \behavior{PRO\-VIDER}) can be used both for leaking 
personal information or for evasion.\footnote{Out-of-the-box 
sandboxes generally have no contacts, which can be leveraged 
to fingerprint these sandboxes.} 
Also, \emph{text} executables represent any %
high-level program that can be interpreted during runtime 
and does not require prior compilation. This includes for 
example Bash shell scripts. \behavior{JAVA\_EXEC} is 
generally used to execute shell scripts during runtime. 
Contrary, \behavior{JAVA\_NATIVE} is used to execute 
ELF binaries. However, \behavior{JAVA\_EXEC} could 
potentially be used to execute ELFs as well. 
\rev{We refer the reader to an earlier version of this work for more details on the relevance of studying sets of API calls~\cite{mirzaei2019andrensemble}. 
For the purpose of this work, we mainly focus on providing a time-line understanding of how malware have evolved.}

Overall, for the 1.2 million apps in our dataset we observe a 
total of 155.7 million methods, out of which about 1.3 million 
are rider methods. %
The average number of methods per app is 121 and the largest 
number of different methods in one single family reaches 16.5 
millions. Overall, each family has on average 1,225 rider methods. %
\subsection{Evolution over Time}
\label{sec:rider:time}

\begin{figure*}[t] 
\centering
\subfloat[Prevalence of sensitive APIs.]{\includegraphics[width=.5\textwidth]{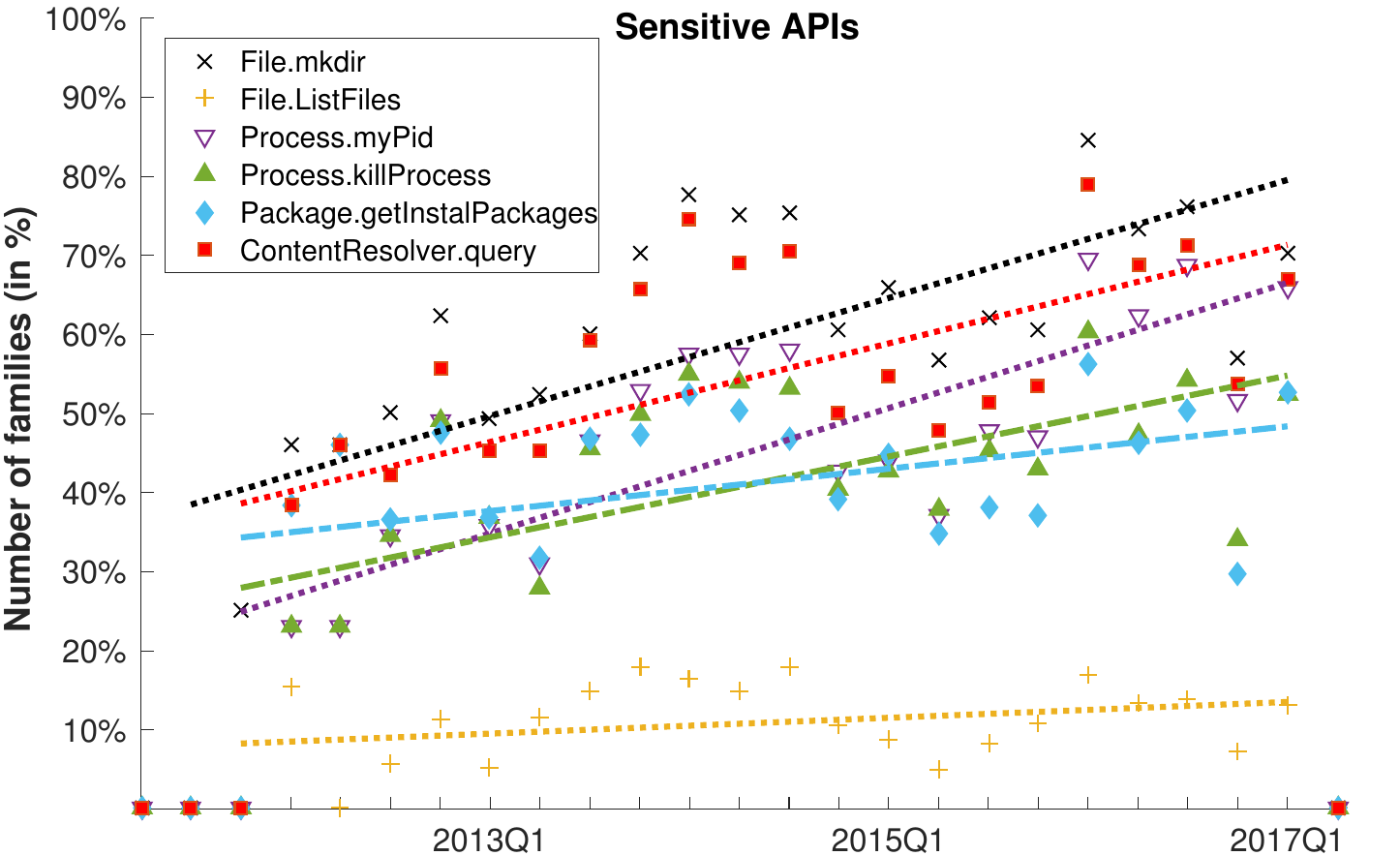}\label{fig:timeline-A}}
\hfill
\subfloat[Prevalence of network comm APIs.]{\includegraphics[width=.5\textwidth]{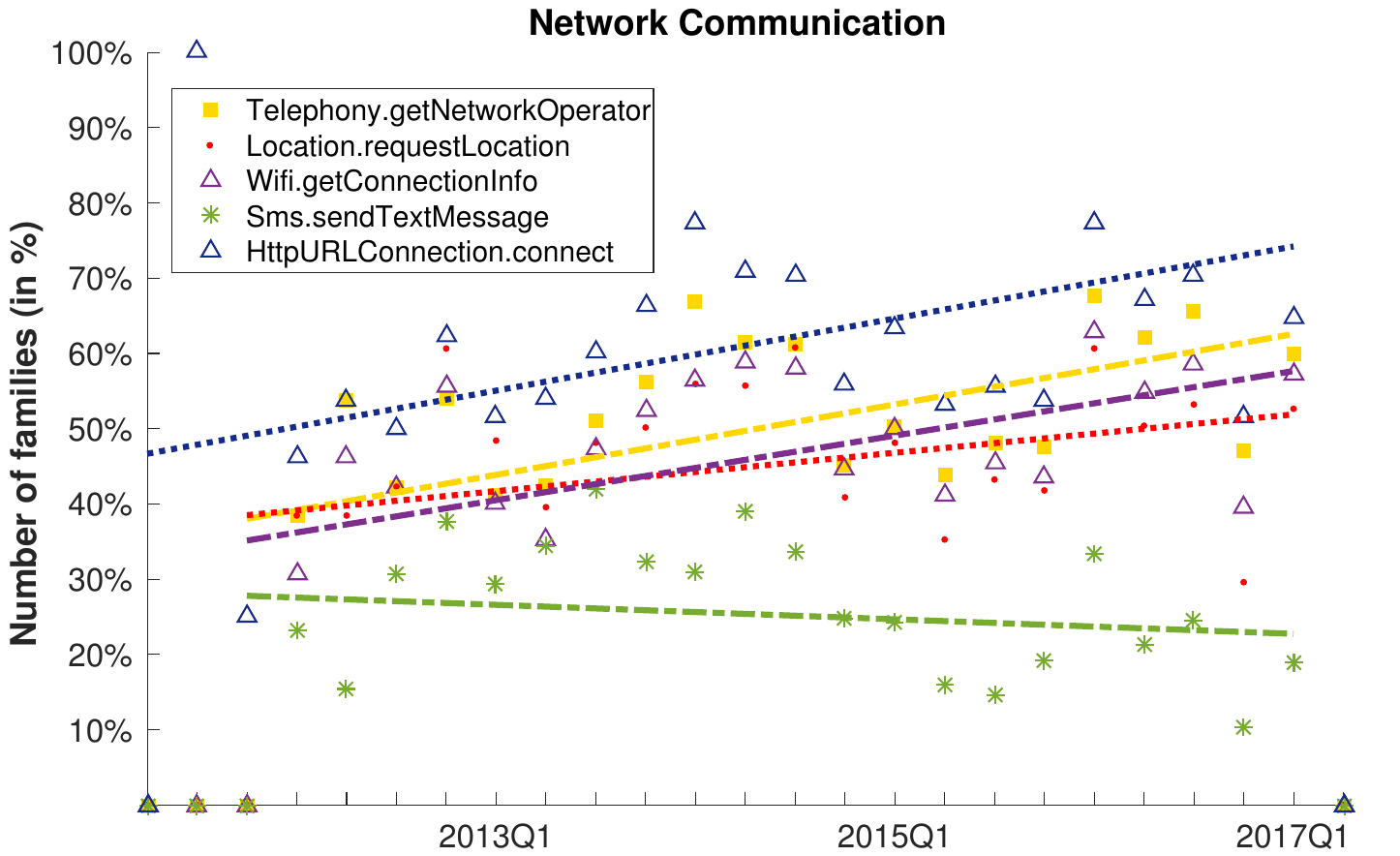}\label{fig:timeline-B}}
\hfill
\subfloat[\rev{Prevalence of obfuscation.}]{\includegraphics[width=.5\textwidth]{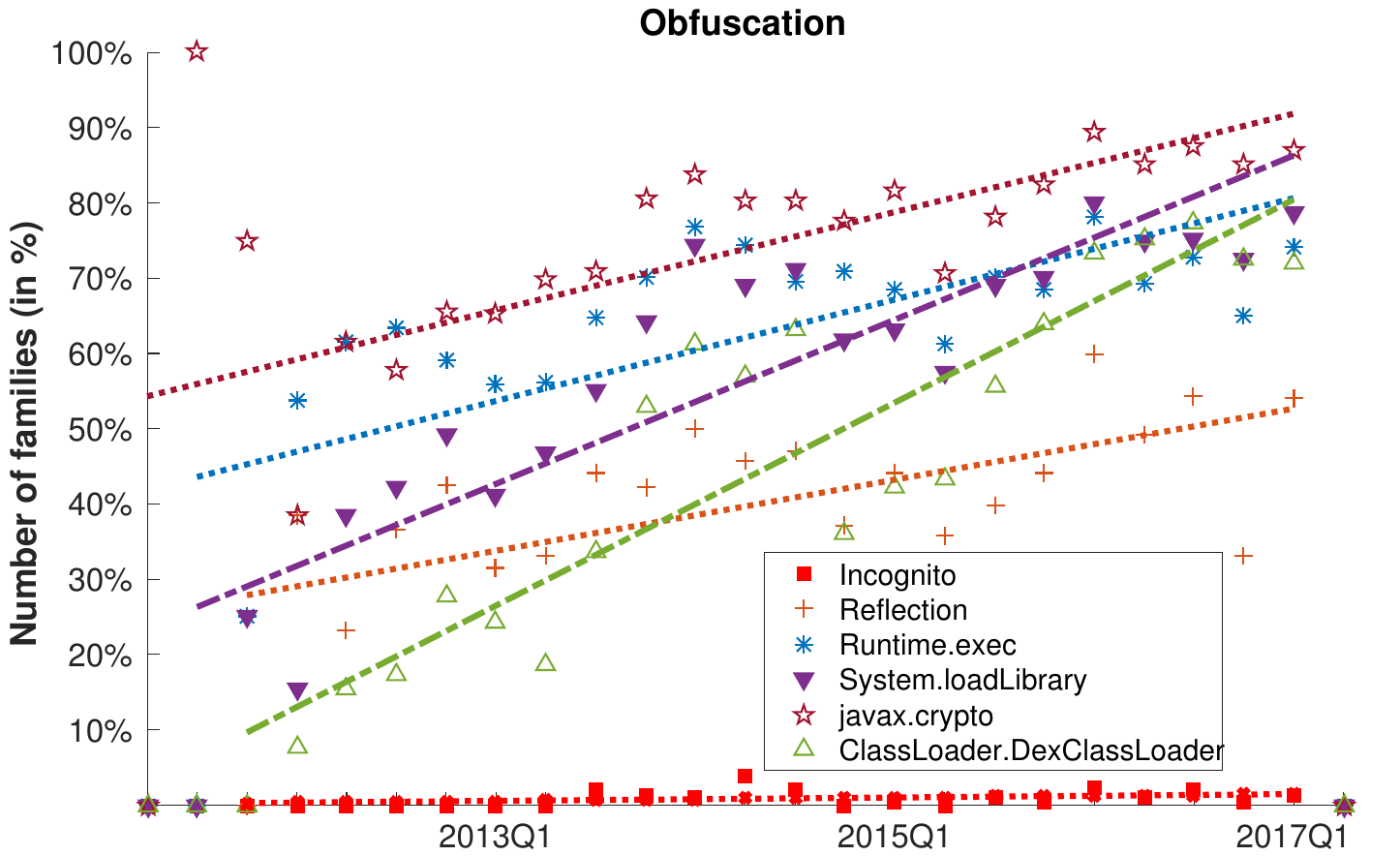}\label{fig:timeline-C}}
\caption{Percentage of families active in each quarter %
where at least 90\% of their members share a feature in common. 
  }
\label{fig:timeline}
\end{figure*}

Malware is a moving target and behaviors drift over time as 
miscreants modify their goals and attempt to avoid detection. 
In this section we measure how malware behavior evolved across several years. 
According to the type of API call, we group behaviors into three categories: 
(i) sensitive APIs, 
(ii) network communication, and 
(iii) obfuscation. 
\rev{Sensitive API calls are permission-protected APIs that are considered to be indicative of malicious functionality.} 
Figure~\ref{fig:timeline} shows behaviors 
associated to families by quarter of a year for each of the categories. 
The graphs represent the proportion of families that exhibit a 
certain capability in a given quarter, showing how families evolve 
over time. 
It is possible to observe that %
the distribution of malware samples per quarter
is not uniform, but there are two spikes in our data, one in Q1 2014 and one in
Q4 2015 (399 and 819 families acting in those quarters,
respectively). 
On average, the number of families observed per quarter is 280. 
Regardless of the presence of these two spikes, when looking 
at how behaviors evolve overall, one can typically observe a 
trend based on how prevalent API calls are across time.
To study this, we plot the best fit to each set of 
API calls using linear regression. 
Note that the cutoff here is applied to the samples of a 
family that were observed during that quarter 
(see Figure~\ref{fig:statsNumberFamilies} for a snapshot of the number of samples and families seen per quarter). 
As samples in a family are scattered throughout time, this timeline 
gives an understanding of how the family evolves, which naturally fits with the notion of variant discussed in~\S\ref{sec:pre}.
\paragraph{Sensitive APIs}

Figure~\ref{fig:timeline-A} shows behaviors related to 
generic actions such as File System (FS) actions or OS-related
APIs. 
FS- and OS-related behaviors are typically found in families 
that attempt to execute an exploit~\cite{aafer2013droidapiminer}. 
These behaviors include the use of API calls such as 
\texttt{Process.killProcess()} or \texttt{Process.myPid()}. 
Also, \behavior{IO} operations such as \texttt{File.mkdir()} 
are used in preparation to the exploitation. 
Other \behavior{IO} operations shown in this category 
(e.g., \texttt{File.ListFiles()}) are commonly used by 
ransomware. 
We present a case study that illustrate 
how ransomware makes use of sensitive APIs in \S\ref{sec:casestudy:ransomeware}. %
Some of these behaviors (such as \texttt{Process.killProcess()}) 
have increased steadily only up to about 55\%. 
Other behaviors have increased more sharply over the last few 
years, such as \texttt{Process.\-myPid()} to all the way to 75\%. 
\paragraph{Network Communication}

Figure~\ref{fig:timeline-B} shows behaviors related to 
network communications in general. 
One of the first takeaways that can be obtained is related to 
the negative trend in the use of the \texttt{Sms\-Manager.\-send\-Text\-Message()} API. 
This API call is usually associated to a common fraud that profits from silently sending premium rate messages. 
As shown in the timeline, this type of malware was popular between 2012 and 2014. 
One factor behind the popularity of this fraud was its simplicity 
(it typically does not require the support of a back-end). 
However, starting from mid 2014 this behavior sees a drop in 
popularity---from about 40\% to 10\% of the families. 
\rev{
Interestingly, we observe that the overall use of the 
\behavior{SMS} category (i.e., macro perspective\footnote{We refer as macro perspective analysis to the measure of common methods using the same cutoffs but without considering the time component.}) %
is lower than in any point of the time line. 
Thus, the level of 
granularity shown when measuring 
rider behaviors in a time-line manner is much more 
precise than when looking at a macro perspective.}
\rev{To put our work in perspective with respect to malware, we compare our findings with those in~\cite{suarez2017droidsieve}. 
Authors show that the prevalence of the SMS category in malware accounts for 47\% of the dataset while it only appears in 2.83\% of the goodware studied. 
This shows that data-driven detection approaches that are trained with a non-representative dataset will perform poorly with recent threats.}

We can also observe that the use 
of the \texttt{Http\-URL\-Connec\-tion.\-connect()} API call has increased over 
the last years. This API call when combined together with those 
related to privacy violations (e.g., \texttt{Content\-Resolver.\-query()})
is commonly used to exfiltrate personal information~\cite{fan2017dapasa}.
This information can then be sold on underground markets 
or used as part of a larger operation~\cite{onaolapo2016happens} 
(see \S\ref{sec:casestudy:advertisment} for a case study 
discussing exfiltration of personal information).
\texttt{Http\-URL\-Connec\-tion.\-connect()} can also be used to retrieve 
new payloads, which is known as update attacks. 
The use of more sophisticated attacks such as those requiring the 
support of a Command \& Control (C\&C) structure indicate a change 
in the way miscreants monetize their creations from the initial 
premium-rate fraud~\cite{suarez2014evolution}. 
This can be attributed in part to the proliferation 
of inexpensive bulletproof servers or robust botnet structures that 
allow campaigns to last longer~\cite{antonakakis2012throw,holz2008measuring}. 

We also observe behaviors that could be aimed at evading dynamic analysis. 
As mentioned earlier, malware often queries certain hardware  
attributes (or sensor values) that are usually set to default in sandboxes. 
This is the case of the values given by 
\texttt{Connec\-tivity.\-getActive\-Network()} or 
\texttt{Wifi\-.get\-Con\-nec\-tion\-Info()} API calls. 
Although the latter is not shown in the figure, both 
increase with a similar trend reaching 70\% and 55\% of the families 
by 2017.

\paragraph{\rev{Obfuscation}}

The use of \behavior{reflection} has increased over the last years from 
slightly over 20\% of the families in 2012 to about 50-60\% in 2016 
and 2017 as shown in Figure~\ref{fig:timeline-C}. 
The use of this feature can be mainly attributed to obfuscation. 
Other forms of obfuscation can be evidenced by looking at the evolution of 
the \behavior{crypto} category, which is one the most prevalent ones. The 
number of families using cryptographic APIs started at 100\% in 2011 and 
dropped to 60\% in the following year. Soon after that, we observe a 
uniform increase reaching 90\% in 2017. This most likely means that back in 
2011 miscreants that started manufacturing malware for Android had 
a high technical expertise. As Android became the platform of choice, 
more actors with different expertise were involved and the use of 
\behavior{crypto} dropped the next years, to become a common feature a few 
years thereafter, perhaps out of necessity to evade malware detection systems.   
Another reason for which cryptography could be used, independent from
obfuscation, could be ransomware. 
Another form of loading Java code during runtime is via the 
\texttt{ClassLoader.DexClassLoader()} API call. Results show that 
the usage of this interface increases over the years 
to 70\% in 2017. To trace the use of \behavior{incognito} apps, 
we recursively looked at all APK and DEX resources 
in the app and analyzed their methods. Common methods originating from 
\behavior{incognito} apps are, however, not prevalent. This means that 
hiding code relevant to the family via incognito apps is not 
popular---note that advanced hiding techniques, such as those in Stegomalware~\cite{suarez2014stegomalware}, 
can be effectively used to evade automated 
systems~\cite{suarez2014stegomalware}. 
 
\rev{
Interestingly, we can observe that the use of \texttt{Sys\-tem.\-load\-Libra\-ry()}, 
which is related to the invocation of native libraries, has increased 
sharply over the years from 25\% in 2011 to 80\% in 2017. 
With a more modest trend we observe that \texttt{Runtime.exec()} is still 
very prevalent nowadays. 
These two APIs are the main entry point for dynamically loading 
non-Java code that is not installed as part of the app. 
The most common executables loaded are \emph{ELF executables} 
and \emph{text executables} respectively. 
As behaviors offloaded 
to these components can not be seen from Dalvik, we deep inspect 
the resources of each app and give an overview of these findings 
in~\S\ref{sec:resources}.%
We give more details about the type of functionality we 
observed in obfuscated malware in~\S\ref{sec:casestudy:new}.
}

Recent work studied the use of obfuscation on goodware on 
the Google Play store~\cite{wermke2018large}. 
Authors showed that less than 25\% of apps 
have been obfuscated by developers. 
Instead, we show that the obfuscation in malware 
is way more prominent. This might explain why the proliferation 
of malware has been so acute over recent 
years---while miscreants can easily process un-obfuscated  
carriers to build new versions of their malware, security 
experts are, more than ever before, confronted with obfuscated riders.
The increasingly prevalent use of \behavior{reflection}, %
of \behavior{native} libraries, and \behavior{scripts} 
indicates that the behaviors that we observe by 
performing static analysis might not constitute the full 
set of actions performed by malware when executed---we refer the reader 
to~\S\ref{sec:discussion} for a discussion on the limitations of our work.
This also means that recent ML-based works in the area of malware 
detection that do not take into account obfuscation 
are most likely modeling the behaviors seen in the 
carriers rather than those belonging to the riders. 
Thus, we argue that there is a strong need for a change of 
paradigm in the malware detection realm. We argue that the community 
should focus efforts on building novel detection techniques 
capable of dealing with obfuscation. %
\section{Analysis of Resources}
\label{sec:resources}

Malware authors often offload payloads from the Dalvik 
executable to make the app look benign to cursory 
inspection~\cite{DroidNative2017,poeplau2014execute}.
We analyze other types of executables that 
are also packed into the APK. In particular, we look at: 
(i) Dalvik, (ii) Text, and (iii) 
ELF executables to provide a cross-layer inspection. 
In summary, we observe that finding common code structures 
in these type of resources is remarkably challenging. 
Overall, we find that only 4\% of the families 
have unencrypted common resources. This is because 
the level of sophistication used to obfuscate these 
resources is more evolved than the one used in 
Dalvik, and can be explained by looking 
at the number of tools (e.g., packers) available to 
obfuscate external resources~\cite{vigna2018malware}. 
The extended version of this paper~\cite{extended} provides further details on the prevalence of common compiled resources and libraries across families.

\section{Case Studies} 
\label{sec:casestudy}

In this section we present two %
case studies to illustrate how differential analysis can be used to analyze and understand rider behaviors. 
In particular, we have selected (i) a case study from a sophisticated long-lasting ransomware campaign, and (ii) two shady advertisement libraries that have infected over 11K 
apps. 
\revb{We also refer the reader to \PublicRepo{} for additional details, including more verbose outputs of our system and additional families.}

\subsection{Ransomware}%
\label{sec:casestudy:ransomeware}

We first study the case of \family{simplocker}, a ransomware that 
has been operating since 2014Q3 and mainly targeted Google Play. 
While there are several ransomware 
families in our dataset such as \family{jisut}, 
\family{slocker}, \family{gepew}, or \family{svpeng} to name a  
few, \family{simplocker} is one of the first confirmed 
file-encrypting malware families targeting 
Android~\cite{simplockerSymantec}. 
The way Android ransomware operated prior to this family made 
file recovery possible without paying the ransom. In particular, 
these early versions attempted to keep user information hostage by 
simply locking their devices but without encrypting the file 
system. 
Technical experts could then bypass the locking mechanism 
using standard forensic tools (e.g., mounting the file system 
from a PC). 

Our dataset accounts for 30 specimens of \family{simplocker} 
with a total of 35,825 distinct methods. Out of those, 
1,166 (3.2\%) methods are common to at least 28 apps. 
We can also find 295 (0.8\%) methods common to all 30 apps. 
We observe the use of the file system (\behavior{IO}
behaviors), the access to personal information (via the 
\behavior{content} provider), and the use of database-related 
libraries (\behavior{DATABASE}). 
Details about the most relevant methods seen in this family are listed in 
Figure~\ref{tab:case:simplocker}. %
When analyzing the common methods found, we can see that 
this family uses the \behavior{DATABASE} library to 
explore DDBB in a common method. %
This library is used to 
explore data returned through a content provider, 
which is used to access data stored by other apps 
such as the contacts app. We also observe that this 
type of ransomware uses its own crypto suit rather than 
relying on standard Java libraries. 
In particular, 
methods in \texttt{com.nisakii.encrypt.*}, 
such as method-662 and method-909 shown in Figure~\ref{tab:case:simplocker},  
are used to encrypt stolen 
files.

Once files are encrypted they are erased 
using the \behavior{java.io.\-File.\-de\-lete()} API call and the \emph{FileProvider} class in the \texttt{Landroid/\-support/\-v4/\-content} library).
in method-1075 (Fig.~\ref{tab:case:simplocker}). 
This library was developed by Google to provide  
new features on earlier Android versions. 

As mentioned before, our system does not make a priori assumptions 
based on the name of the package or its provenance. 
This is simply because ``legitimate libraries'' can be used 
with a malicious intent\footnote{Recall that the term 
\textit{legitimate libraries} refers to packages that 
are prevalently used in goodware or have been developed by a 
trusted party.}. 
This is precisely what happens with method-1075 ---while 
the library is built by Google and widely used in goodware, 
\family{simplocker} heavily relies on it for malicious purposes.

\begin{figure}
\centering
\lstset{basicstyle=\footnotesize\ttfamily}
\begin{lstlisting} 
    Method-662:
      Seen in: 30 apps (out of 30)
      Class Name: Lcom/nisakii/encrypt/msg/
                    EncryptFragment$EncProcess$2$1;
      Method name: <init>
      Behaviors: {ANDROID, CONTENT}
    
    Method-909:
      Seen in: 29 apps (out of 30)
      Class Name: Lcom/nisakii/encrypt/msg/
                    RegistrationActivity;
      Method name: onBackPressed
      Behaviors: {ANDROID, CONTENT}
    
    Method-1057:
      Seen in: 28 apps (out of 30)
      Class Name: Lnet/sqlcipher/CursorWindow;
      Method name: onAllReferencesReleased
      Behaviors: {ANDROID, DATABASE}
    
    Method-1075:
      Seen in: 29 apps (out of 30)
      Class Name: Landroid/support/v4/content/
                    FileProvider;
      Method name: delete
      Behaviors: {ANDROID, SUPPORT, IO}
\end{lstlisting}
\caption{Excerpt of riders %
for \family{simplocker}.}\label{tab:case:simplocker}
\end{figure}

\subsection{Adware}%
\label{sec:casestudy:advertisment}

We next present the case of two adware families named \family{utchi} 
and \family{lockad}. This form of \emph{fraud} typically monetize 
personal information to deliver targeted advertisement campaigns. 
While the campaign delivered by the former family has been operating 
for over three years and it is one of the most viral families, the latter 
is characterized by its novelty and stealthiness, and displays a clear 
distinction in the complexity of the malware evolution.

\family{utchi} is a family named after a shady advertisement 
library that leaks the user's personal information after being 
embedded into the infected app. 
The library has been piggybacked into over 13K apps distributed 
throughout different markets such as AppChina, Anzhi, and Google Play. 
This family %
mostly operated between the end 
of 2015 and early 2016, although the campaign had been running 
for almost three years since the end of 2013.
Our analysis found 27 methods 
with sensitive behaviors (cf. Section~\ref{sec:rider:behavior})
common to more than 12K apps. %
Among others, these behaviors include \behavior{network} activity, 
access to \behavior{content} provider, access to 
unique serial numbers (via the \behavior{telephony} manager), and 
the use of \behavior{reflection}. 
Similar behaviors can be seen in other data-hungry advertisement networks  
such as those observed in 
\family{leadbolt}, 
\family{adwo}, 
\family{Kugou}, or 
\family{youmi}.

Similarly, \family{lockad} piggybacks some libraries that are 
used to exfiltrate personal information from the user to later 
display unsolicited advertisements. Some of the services 
that are loaded as part of the infected app are: 
\texttt{com.dotc.ime.ad.service.AdService} or 
\texttt{mobi.wifi.adlibrary.AdPreloadingService}. 
To avoid detection and hinder static analysis, samples in 
this family obfuscate certain core components of the 
embedded library. 
For instance, the library unpacks configuration parameters 
from an encrypted asset-file called `cleandata' 
as shown in Figure~\ref{tab:case:lockad} (method-345). 
These parameters 
are later used to decrypt additional content fetched from the 
Internet. 
Method-4670 contains the decryption routine that 
uses standard AES decryption in CBC mode and with PKCS5 Padding. 
The routines displayed in this figure have been reverse-engineered 
and method names (e.g., \texttt{make\_md5})  
have been renamed to better illustrate the behavior of this method. 
Finally, we can also observe in this family methods that provide 
support to run Text Executables. When running a dynamic 
analysis of one of the samples\footnote{We have used a dynamic analysis system called CopperDroid~\cite{tam2017evolution}.}, we could corroborate that 
the APIs seen attempted to invoke several processes 
(e.g., \texttt{/proc/*/cmdline}) to run the executables.

\begin{figure}[bt]
\centering
\lstset{basicstyle=\footnotesize\ttfamily}
\begin{lstlisting} 
Method-345: Seen in 6 apps (out of 7)
 Class Name: Lrk; Method name: a
 Behaviors: {NET}; Routine: 
   stream = pContext.getAssets().open("cleandata");
   a = new JSONObject(rg.a("hwiHQwgVw2I", stream));

Method-4670: Seen in 7 apps (out of 7)
 Class Name: Lrg; Method name: a
 Behaviors: {CRYPTO}; Routine: 
  String a(String key, InputStream param){
    Cipher c = Cipher.getInstance("AES/CBC/PKCS5");
    byte[] md5 = make_md5(key.getBytes("UTF-8"));
    c.init(2, new SecretKeySpec(md5, "AES"), IV);
    CipherInputStream is;
    is = new CipherInputStream(param, c);
     [ Calls to CipherInputStream byte by byte ]}
\end{lstlisting}
\caption{Excerpt of riders %
for \family{lockad}.}\label{tab:case:lockad}
\end{figure}

Apart from leaking personal info., both families also use 
\behavior{reflection} to dynamically load new functionality.

\subsection{First Seen 2017Q1}%
\label{sec:casestudy:new}

We now present the case of a family called \family{hiddenap} 
that was first seen in early 2017 and soon after accounted 
for 83 samples in our dataset. 
\family{hiddenap} is one of the largest families seen in 
2017\footnote{Together with a family called \family{ggsot}.}. 
Apps in this family are mainly distributed through alternative 
markets and all samples in our dataset have been obtained from 
one of the largest Chinese alternative market (i.e., the Anzhi market). 
This family is fairly basic and it only has 17 methods common 
to all apps. These methods exhibit behaviors mainly related 
to \behavior{IO} operations together with other standard 
actions from the Android framework such as the \behavior{content} 
provider. %

Once the device is infected, the malware runs an \emph{update 
attack} in a method called \texttt{com.secneo.guard.Util.checkUpdate()} 
(method-17 in Figure~\ref{tab:case:hiddenap}). 
It then attempts to drop additional apps and install them with 
the support of some native libraries called \texttt{libsecexe.so}, 
\texttt{libsecpreload.so}, and \texttt{SmartRuler.so} 
that are embedded into the app. 
The last two libraries have been seen together with 
apps that are packed using a known service called Bangcle%
\footnote{https://www.bangcle.com/ (in Chinese).}~\cite{zhang2015dexhunter}. 
The third library most likely contain an exploit that would grant 
root privileges to the malware. 
All native libraries are compiled both for x86 and ARM processors. 
Before loading the library, the malware first checks which is 
the right architecture of the device with 
\texttt{Lcom/\-secneo/\-guard/\-Util.\-check\-X86()} 
(method-7) 
using standard API call such as \behavior{File.exists()} or 
\texttt{System.getProperty()}. 
\begin{figure}
\lstset{basicstyle=\footnotesize\ttfamily}
\begin{lstlisting} 
    Method-7: 
      Seen in: 83 apps (out of 83)
      Method name: checkX86
      Behaviors: {ANDROID, CONTENT, IO}
        
    Method-17: 
      Seen in: 83 apps (out of 70)
      Method name: checkUpdate
      Behaviors: {ANDROID, CONTENT, IO}
\end{lstlisting}
\caption{Excerpt of common methods %
for \family{hiddenap}.}\label{tab:case:hiddenap}
\end{figure}

Even though this family is using a packer to obfuscate parts 
of the code, the hook inserted in the Java part has meaningful 
method names that convey very accurately what the malware does. 
Considering that the app is obfuscated 
using an online packer, we can conclude that this 
miscreant had a limited technical background.

\section{Discussion}
\label{sec:discussion}

In this section, we first discuss a number of limitations 
of our study. %
We then highlight the most important findings observed and 
discuss their implications for future research.

\subsection{Limitations}
\label{sec:discussion:limitations}

A sensible goal for a malware developer is to 
obfuscate the rider or offload it remotely. 
We next discuss the challenges behind 
these threats and the main 
limitation. %
\paragraph{Obfuscation}

Our study inherits the limitations of static analysis and 
thus can unavoidably miss obfuscated riders. In fact, we 
have observed that the use of cryptographic APIs 
has increased significantly over the years. 
This problem is the scope of our future work as we explain next. 
Even when specimens rely on obfuscation, due to the nature 
Android they nonetheless require a trigger 
that would deobfuscate the payload. We can isolate these triggers 
using differential analysis as done in this work. This can aid dynamic 
analysis techniques to fuzz only those classes (and methods) where 
the hook to the obfuscated payload rests. Dynamic behaviors emanating 
from those payloads can then be used to extend the set of behaviors 
seen statically. 

The underlying technique that we use to compute differential analysis 
assumes that piggybacked classes respect the morphology of their 
code (in terms of CFG). There are advanced obfuscation techniques 
such as polymorphic and metamorphic malware that could alter the 
structure of the code (including the CFG of their methods). 
Furthermore, recent work shows that it is feasible to use 
stegomalware to systematically add dynamic code~\cite{suarez2014stegomalware}. 
This would render differential analysis useless. However, to the 
best of our knowledge, there is no evidence in the wild that would 
indicate that this type of obfuscation is used in Android malware at large. %
\paragraph{Update Attacks}

In update attacks, the rider is loaded at runtime~\cite{poeplau2014execute}. 
Typically, the payload is stored in a remote host and 
retrieved after the app is executed~\cite{DroidNative2017}. Unless the rider 
is stored in plain text within the resources of an app, 
our study is vulnerable to this attack. 
We could overcome this limitation in a similar way 
as in the case of obfuscation---using dynamic 
analysis. For local update attacks, we recursively 
inspect every resource to find \textit{incognito} apps. 
We append the methods of those apps to the methods of 
the main executable before running our differential 
analysis system.

\paragraph{Notion of Family}

The way in which differential analysis is used in 
this paper requires a precise accounting of the 
members in a family. To do so, we rely on 
Euphony~\cite{hurier2017euphony} which in turn 
leverages threat intelligence shared from multiple 
AV vendors. Unifying diversified AV labels is a challenging 
process that might be subject to misclassifications. 
This is because Euphony is forced to make decisions 
based on information given by AVs, whose family definitions 
might disagree with each other. Unifying labels is thus
prone to error (especially with very recent families). 
Furthermore, the morphing nature of malware renders the notion 
of family incomplete and makes differential analysis dependent 
on the variants. 

In our work, we overcome these challenges by introducing %
a relaxed cutoff that can flexibly be configured. 
For the case of API-based behaviors (\S\ref{sec:rider}), 
we set the threshold to 90\% rather than 100\% to minimize 
the impact of potential misclassifications in Euphony. 
The selection of this threshold was motivated by 
the performance reported in~\cite{hurier2017euphony}. 
In this paper, we showed that 
grouping samples chronologically provides a more granular way 
to understand how variants behave and, ultimately, 
how families evolve. 

The cutoff chosen can cover a wide range of variants 
when combined with a chronological grouping. 
In any case, one could set even lower thresholds to fine-tune the granularity of the variants observed. However, this could risk the inclusion of code fragments coming from the carriers. 
This is because different goodware can import the same libraries as discussed in~\cite{li2016investigation}. 
One option could be to `white-list' those libraries and 
remove known software components before applying differential analysis. 
Along these lines, Google has recently proposed the use of 
what they call \emph{functional peers} to set as `normal' 
behaviors that are often seen in known goodware of the same 
category (peers)~\cite{googleblogFunctionalPeers}. 
However, in our work we choose not to do this. The main reason 
behind this is that legitimate libraries can also be used with 
  a malicious intent (see for example the case study described in \S\ref{sec:casestudy:ransomeware}) %
For our purposes, we consider that keeping a threshold %
relatively high (i.e., above 90\%) is enough to avoid including 
code fragments from the carriers. 

In this work we assume that if a method appears in a large 
portion of samples in a family, it can be considered harmful 
or it could potentially be used maliciously by most of the members of a 
family. 
As part of our future work, we are planning to \textit{taint} all common methods that appear frequently in top ranked apps in Google Play. 
This way, we could study how common libraries are invoked by malware. 
We also want to leverage on existing knowledge about piggybacked pairs of raiders and carriers to also taint 
methods that appear to be common~\cite{chen2015finding,li2017understanding}. 
In this case, common methods would reveal those software fragments 
that belong to the carrier as opposed to what we do in our paper. 
Tainted methods could also be used to elaborate on the aforementioned 
concept of \emph{functional peers} to provide stronger guarantees 
of the software provenance in repackaged malware. This information can help to provide a notion of risk, where 
behaviors that appear mostly in riders and are never seen in 
goodware should be considered highly risky and vice versa.

\paragraph{Studied APIs}

In this paper we reported our 
findings after analyzing %
the most frequent set of APIs used by riders as well as those considered to be the most indicative of maliciousness as described in~\S\ref{sec:rider:time}. 
We acknowledge that the set of API calls falling into one of the generic 
categories can change over time. However, we note that 
these APIs are categorized based on the package name 
(see~\S\ref{sec:rider:behavior}). 
Thus, when a new API call is added, for instance, to the 
\behavior{TELEPHONY} category, we can guarantee that the 
API call is related to the Telephony module of the Android 
OS. 
When looking at APIs individually, we have carefully 
investigated the official documentation of every single 
API call relevant to our study. We have observed that 
all API calls discussed in this paper were introduced as 
part of the foundations of the Android OS framework 
(between API level 1 and 4\footnote{Added between Android 
1.0 in 2018 and Android Donut in 2009.}) and have not been 
deprecated at time of writing. We have observed in our static traces, 
however, one API call (i.e., \texttt{Activity\-Manager.\-get\-Running\-Services()})
that has been eventually used by riders and that was 
deprecated in API level 26 (Oreo 8-0, August 2017). 
As part of our future work, we would like to explore 
the implications that deprecated API calls have for 
malware developers.

\subsection{Key Findings} %
\label{sec:discussion:findings}
While our study presents the limitations as discussed above, 
we observe a large number of apps displaying common, and more 
importantly, sensitive behaviors. Our findings constitute a 
large-scale longitudinal measurement of malice in the Android 
ecosystem. We next summarize the key takeaways of our work and discuss 
their implications to research.

\paragraph{Threat evolution}

Our results show that certain threats have evolved rapidly over 
the last years. For example, premium-rate frauds that 
were seen in about 40\% of the families in 2013 and dropped to 
10\% in late 2016. 
On the contrary, the use of native support  
has increased sharply from 15\% in 2011 to 80\% in 2017. 
\rev{We have also noted that that looking at large families alone (without considering notion of variants) can provide misleading interpretations of the evolution of malware.
We have further shown that the time component is paramount to disambiguate the notion of variant. 
Our findings show that works such as~\cite{fan2018android} are hard to deploy in real-world settings.} 

This shows the importance of a time-line evaluation when 
developing new malware detection approaches, together with 
the need for research outcomes reporting results on samples 
with features tailored to the type of threat faced in each period. 
Recent work~\cite{drebin,suarez2017droidsieve} 
neither report time-lined results, nor use features from native libraries. 
These two items should constitute a guideline for future research 
in the area of malware detection. 
Authors in~\cite{mariconti2016mamadroid} investigated the evolution of 
malware detection over time up until 2016, but did not look at 
how samples change. 
\rev{Finally, more a recent work has studied ways to eliminate experimental bias in malware detection~\cite{pendlebury2019tesseract}. 
Authors have shown that models trained with machine-learning algorithms should be aware of the temporal axis to provide reliable results in real-world setting. 
In particular, authors emphasize the importance of having a temporal goodware to malware window consistency. 
However, this still remains an open research question and concrete steps as of how to achieve a window consistency are needed. 
The methodology used in our paper can be used to better understand malware variants in a temporal-manner to address this open issue. 
}

\paragraph{Evidence of obfuscation}

A large scale investigation of the use of 
obfuscation in Google Play have recently shown that 
only 24.9\% of the apps are obfuscated~\cite{wermke2018large}. 
In this work we look at evidence of obfuscation
among riders. In particular, we study the usage of crypto 
libraries and three different forms of dynamic code 
execution: native code, Dalvik load, and script execution. 
We show that all forms of obfuscation are increasingly 
more popular in malware, with the usage of cryptography 
present in 90\% of the families in 2017. When putting 
this in perspective with respect to legitimate 
apps~\cite{wermke2018large,vigna2018malware}, 
we highlight a sharp increase in the use of these techniques. 
Discussions about the attribution of certain behaviors 
such as the use of obfuscation to repackaged malware 
have been recurrent in literature over the last few 
years~\cite{lindorfer2014andrubis}. Our findings 
suggest that malware developers are ahead 
of legitimate ones. 

To the best of our knowledge, there are 
few malware detection systems capable of dealing with 
these forms of obfuscation. For the case of reflection, 
the authors of~\cite{rasthofer2016harvesting} proposed {\sc HARVESTER}, 
a system that can resolve the targets of encoded reflective 
method calls. For the case of incognito apps, authors 
in~\cite{suarez2017droidsieve} look at inconsistencies left 
by this type of obfuscated malware. While these   
approaches can deal with certain types of obfuscated 
malware, they are vulnerable to motivated adversaries. 
For instance, {\sc HARVESTER} can not deal with 
static backward slicing attacks. 

Dynamic analysis constitutes the next line of defense 
against obfuscation~\cite{tam2017evolution}. However, 
we have also observed that evasion is not only becoming 
more popular, but also more diverse. The research 
community has recently positioned that evasion attacks can be addressed 
with static analysis~\cite{fratantonio2016triggerscope}---triggers 
can be first identified using symbolic execution and a smart stimulation 
strategy can then be devised. One major challenge here 
arises from the combination of obfuscation and evasion 
attacks. For instance, an adversary can use opaque 
predicates to hide the decryption routine of the 
malware to defeat both static and dynamic analysis. 

\paragraph{Standalone malware}

We do not make claims about the amount of standalone malware 
(i.e., malware that does not take advantage of repackaging) in 
the wild but we can report an estimate as depicted in our 
dataset. While we found that 25 families (542 samples) out 
of 1.2K+ families (1,282,022 million samples) could potentially 
be standalone malware, we also discarded all families with less 
than 7 samples per family from the original set of 3.2K+ families
(1,299,109 samples). 
This was due to the way differential analysis works, which 
requires a critical mass of samples. Given that standalone 
malware tends to have a small number of samples per family, 
one could assume that most of the samples discarded are 
standalone malware. If this holds true, a fair approximation 
of the number of standalone malware would then be about 
2K families and 17K samples (62.5\% of the families, but only 
1.36\% of the samples). 

On the other hand, our dataset only contains samples that have 
been labeled into families by Euphony~\cite{hurier2017euphony}. 
Unlabeled samples are known as \emph{Singletons}, and there 
are about 200K of these in the AndroZoo~\cite{allix2016androzoo} 
dataset as of the day we queried it. 
If we were to assume that all singletons are standalone malware, 
we will then be looking at figures of approximately 13\%. 

While we estimate that standalone malware could range between 
1.36\% and 13\% of the total malware in the wild, the authors of~\cite{wei2017deep} 
report that 35\% of the samples in their study are standalone 
malware. They analyze roughly 405 samples, sampled from a larger 
dataset. %
Interestingly, some 
of the families that are flagged as standalone contain a large 
number of samples (e.g., \family{lotoor} and \family{opfake} 
with 1.9K and 1.2K samples respectively), which seems unlikely. 
\section{Related Work}
\label{sec:related}

A number of papers analyzed Android malware 
over the last years~\cite{aafer2013droidapiminer,lindorfer2014andrubis}. 
One of the key aspects to consider when systematizing the analysis 
of malware is properly curating the dataset to remove potential 
noise from samples. 
Works in the area of malware network analysis 
have recently shown that this process is of paramount importance~\cite{lever2017lustrum}. 
In the Android realm, this is especially challenging due to the 
proliferation of repackaging. 
We tackle this challenge by using of differential analysis, which is based on static analysis. 
Static analysis has been used in the past to systematize the study of the Android app ecosystem~\cite{enck2011study}.
However, up until now this was not used to study malware at large.

There have been several works 
looking at piggybacked malware in the last few 
years~\cite{zhou2012detecting,zhou2013fast,chen2015finding,DroidNative2017}. 
In the case of MassVet~\cite{chen2015finding}, 
authors propose a similar methodology than the one we propose to find commonality among apps. 
However, their focus is on the detection of repackaging via similarities in the GUI. 
In DroidNative~\cite{DroidNative2017}, the authors look at the CFG of native code to distinguish between goodware and malware.  
Instead, we mine common code structures and measure 
the prevalence of API-call usage. %
Furthermore, our dataset of malware is about one order of magnitude larger than the one used in MassVet and about three order of magnitude larger than in DroidNative. 

Li et al.~\cite{li2017understanding} 
propose a system to detect piggybacked apps. They also investigate 
behaviors seen in riders, however a key difference with our work 
is that they compare pairs of piggyback-original apps 
individually rather than providing a per-family overview. 
The scope of their work is limited to 950 pairs as opposed to our work.
The main advantage behind a per-family longitudinal measurement 
is that findings here provide a holistic overview of the prevalence and evolution of malice. %

Recent works in the area have proposed the use of common 
libraries to both locate malicious packages in piggybacked 
malware~\cite{li2017automatically} and to create white-lists 
of Android libraries used in goodware~\cite{li2016investigation}. 
In these two approaches, they leverage the library name to 
build a package dependency graph and measure the similarity 
between package names.
\cite{li2017understanding} also uses package name matching 
to infer the ground truth. In our work we choose not to 
rely on the package names as these can be easily obfuscated. 
Instead, we look at the CFG of different code units (methods). 
One major advantage of focusing on the internal structures of 
code is that it provides an improved resistance against obfuscation.

A recent paper by Wang et al. analyzes a 2017 snapshot of the apps available on 16 Chinese Android markets~\cite{wang2018beyond}. 
They show that malware is relatively commonplace on these platforms, and that repackaging is less common by what we observed in our longitudinal measurement in this paper.
Finally, other more recent works have analyzed the evolution 
of Android by looking at permission requests~\cite{calciati2017apps}.  
Similar to the case of package names, the granularity obtained 
from permissions is not as precise as API-annotated CFG.

\section{Conclusions}
\label{sec:conclusions}

In this paper, we presented a systematic 
study of the evolution of rider behaviors in 
the Android malware ecosystem. We addressed  
the challenge of analyzing repackaged malware 
by using differential analysis. 
Our study provides a cross-layer perspective 
that inspects the prevalence of sensitive behaviors
in different executables, including native libraries. 
Our analysis on over 1.2 million samples that span 
over a long period of time showed that malware threats on Android 
have evolved rapidly, and evidences the importance 
of developing anti-malware systems that are resilient 
to such changes. 
This means that automated approaches relying on machine-learning 
should come together with a carefully crafted 
feature engineering process, trained on datasets that are as 
recent as possible and well balanced. 
We have further discussed what our findings mean for Android 
malware detection research, highlighting other areas that 
need special attention by the research community.

\section*{Acknowledgments}
We thank the authors of Dendroid~\cite{suarez2014dendroid}, 
Androzoo~\cite{allix2016androzoo}, and Euphony~\cite{hurier2017euphony} for releasing their work and letting use their service for research. 
We would also like to thank the anonymous reviewers for their comments.

\bibliographystyle{IEEEtran}
%\bibliography{Bibliography}

%
%
%
%
%

%

%

\end{document}